\documentclass[]{aa}
\usepackage{amssymb}
\usepackage{epsfig}
\usepackage{times}
\begin{document}
\thesaurus{02(11.04.2, 11.09.1, 12.04.1, 13.25.2)}
\title{Dark Matter in the Dwarf Galaxy NGC\,247}
\author{Marcus Str\"assle\inst{1}, Marius Huser\inst{1,2}, 
Philippe Jetzer\inst{1,2} and Francesco De Paolis\inst{3}}
\offprints{marcus@physik.unizh.ch}
\institute{
Institut f\"ur Theoretische Physik
           der Universit\"at Z\"urich,
           Winterthurerstrasse 190,
           CH-8057 Z\"urich
\and
Paul Scherrer Institut, Laboratory for Astrophysics,
CH-5232 Villigen PSI
\and
Department of Physics and INFN, University of Lecce, CP 193, I-73100 Lecce, 
Italy
}

\date{Received 25 January 1999 / Accepted 24 June 1999}

\maketitle
\markboth{Dark Matter in the Dwarf Galaxy NGC\,247}{Dark Matter in the
Dwarf Galaxy NGC\,247}
\begin{abstract}
Dwarf galaxies are dominated by dark matter even in the innermost
regions and, therefore, provide excellent probes for the investigation
of dark halos.
To that purpose, we analyse ROSAT PSPC-data of the dwarf galaxy NGC\,247. We focus in
particular on the diffuse X-ray emission in the $1/4\,$keV band.
Assuming an isothermal density profile, we find that the
mass of the hot emitting gas is about
$10^8\,{\rm M_{\odot}}$, corresponding to $\lesssim 0.5\%$ of
the total dynamical mass of the galaxy.
The total mass of NGC\,247, as derived from the X-ray data agrees
quite well with the value obtained from the measured rotation curve (Burlak
\cite{burlak}).

The X-ray profile in the $3/4\,$keV and $1.5\,$keV band shows an excess
at a radial distance of about $15\,$ arcmin from the center. Such a ``hump''
in the radial X-ray profile can be explained by the presence of a cluster
of young low mass stars or brown dwarfs. Therefore, NGC\,247 offers the
possibility to observe the formation of a halo of MACHOs.
\keywords{Galaxies: dwarf -- Galaxies: individual: NGC\,247 -- dark matter -- X-rays: galaxies}
\end{abstract}

\section{Introduction}
Dwarf galaxies (hereafter DGs) differ from normal spiral galaxies in
many of their properties. While the rotation curves of spirals tend to be flat
after the  central rising, rotation curves of  DGs continue to rise out to the
last measured point, although, recently in DDO\,154 it has been observed
that the rotation curve is declining in the outer parts (Carignan \& Purton \cite{cap}). Moreover, DGs tend to have a disk gas content (measured
through the $21$\,cm line) higher with respect to that of spirals. 
In fact, within the radius $D_{25}/2$, defined to be the radius
at which the surface brightness reaches $25^m$ arcsec$^{-2}$, the ratio between
the mass of the gas $M_g$ and the mass of the visible stars 
$M_*$ is $M_g/M_*\simeq 0.45$
and tends to increase to $M_g/M_*\simeq 0.7$ at the radius $LP_{HI}$ of the last
HI profile point (Burlak \cite{burlak}). 
The contribution of the dark matter $M_d$ in DGs within 
$3r_0$ ($r_0$ being the scale height of the stellar 
disk as obtained from surface photometry) is on average 
$65\%$ of the total mass $M_t$, which is nearly a factor two greater than for
normal spiral galaxies. It is interesting to note that among the 13 DGs
analysed by Burlak (\cite{burlak}), the 8 DGs dimmer than $M_B=-17^m$
have on average $M_d/M_t\simeq 0.7$ and $M_d/M_*\simeq 3.9$ while the 5 DGs
with absolute magnitude in the range $-17^m>M_B>-18^m$ have
$M_d/M_t\simeq 0.56$ and $M_d/M_*\simeq 1.5$.  
For comparison, we note that in the case of normal spiral galaxies 
$M_d/M_t\simeq 0.38$ and $M_d/M_*\simeq 0.76$.

Although observations of dwarf galaxies are rather difficult, 
we can learn a lot from them. 
Dwarf galaxies are the most preferred objects for studies of the distribution
of the dark matter,
since the dynamical contribution of their dark halos
dominates even in their central regions. 

DGs have been already precious objects, since they 
allowed to show that their dark
matter constituent cannot be hot dark matter (neutrinos) 
(Tremaine \& Gunn \cite{trg}, Faber \& Lin \cite{fablin},
Gerhard \& Spergel \cite{gersp}).   
DGs put also constraints on cold dark
matter candidates (WIMPs).
Recent studies seem to point out that there is a discrepancy between the 
computed (through N-body simulations) rotation curves for dwarf galaxies
assuming an halo of cold dark matter and the measured curves (Moore
\cite{moore}, Navarro et al. \cite{navarro}, Burkert \& Silk
\cite{burkert}).
Since DGs are completely dominated by dark 
matter on scales larger than a few kiloparsecs (Carignan \& Freeman \cite{cf}),
one can use them to 
investigate the inner structure of DG 
dark halos with very little ambiguity about the contribution from the 
luminous matter and the resulting uncertainties in the disk 
mass/luminosity ratio (M/L).
Only about a dozen rotation curves of dwarf galaxies have been measured, 
nevertheless a trend clearly emerges: 
the rotational velocities rise over most of the 
observed region, which spans several times the optical scale lengths,
but still lies within the core radius of the mass distribution.
Rotation curves of dwarf galaxies  
are, therefore, well described 
by an isothermal density law.

The diffuse X-ray emission has not yet been analysed for many DGs.
Such observations yield important information about the
dynamical mass distribution in elliptical galaxies.
It is, therefore, natural to investigate if also DGs have a diffuse X-ray 
emission and if it can give us useful informations about their total
mass. Of course, we are aware that this is a difficult task.
Beside a proper treatment of the observational background, we must be able
to disentangle the X-ray emission due to the hot halo gas from
the contribution of point sources like low mass X-ray binaries or supernova
remnants, which is demanding even for high mass ellipticals.
In DGs the problem is even more severe, since we expect a weaker diffuse
X-ray emission due to the less deep gravitational potential as compared 
to that of elliptical galaxies.
This implies a lower diffuse gas temperature $T_g$ and, therefore, less
emission in the X-ray band. 

Five DGs have been observed in the $0.11-2.04$\,keV
band with the ROSAT PSPC. Among them, we analysed the three members
of the Sculptor group NGC\,247, NGC\,300 and NGC\,55. In this paper we
present our results for NGC\,247, for which we have the best data.

Unfortunately NGC\,247 is not the best example of a dwarf galaxy.
It has a huge gaseous disk and the dark matter begins to dominate over 
the stellar and gaseous disk only beyond 8\,kpc
(see Fig.~1 in Burlak \cite{burlak}). However, its halo parameters seem
to be typical for the whole class (Burlak \cite{burlak}). In particular, the
ratio between the halo mass and the total mass within the last measured
HI profile point could be as high as $M_h/M_t \simeq 0.7$, which is significantly
above the value for spiral galaxies.

The paper is organised as follows:
In Sects.~2 and 3 we discuss the data analysis procedure and the model
assumptions.
In Sect.~4 the results of the spectral analysis are
shown, while in Sect.~5 we determine the mass of the X-ray emitting gas 
and the dynamical mass of NGC\,247. 
We close the paper with our conclusions in Sect.~6.

\section{Data Analysis}

The data analysis is based upon the standard model for DGs of 
Burlak (\cite{burlak}), which 
includes three components: a stellar disk, a gaseous disk and a dark halo.
In addition, we consider a fourth component, the hot diffuse gas, 
which turns out to be not very important with respect to the 
total mass, but is relevant in our
analysis, since it is the source of the diffuse X-ray emission.
This component is assumed to form a halo 
around the galactic center, which can be described by a $\beta$-model 
(Canizares et al. \cite{canizares}) with central density
$\rho_{\circ}$ and core radius $a_g$ different from the corresponding values 
of the dark halo hence,
\begin{equation}
\rho_{hg} = \rho_{\circ}\left(1+\left(\frac{r}{a_g}\right)^2\right)^{-\frac{3}{2}\beta}\,.
\label{betamodell}
\end{equation} 

The X-ray analysis does only depend on the model of the hot emitting gas.
We will, therefore, not make any further assumptions about the remaining 
components. The hot, diffuse gas component is taken to be in hydrostatic
equilibrium. For spherical symmetry the dynamical mass $M(r)$ 
within $r$ is then given by
\begin{equation}
M(r)=-\frac{kT_{hg}(r)r}{\mu m_p G}\left(\frac{d \log \rho_{hg}(r)}{d\log r} + 
\frac{d\log T_{hg}(r)}{d\log r}\right)\,,
\label{M(r)}
\end{equation}
where $T_{hg}$ denotes the hot gas temperature, $\rho_{hg}$ its density,
$m_p$ the proton mass and $\mu$ the mean atomic weight.§

For the data analysis and reduction we follow the method
proposed by Snowden \& Pietsch (\cite{snowden}) using Snowden's software 
package especially written
for the analysis of ROSAT PSPC data of extended objects and the
diffuse X-ray background (Snowden \& Pietsch \cite{snowden}, Snowden et al. 
\cite{contamination}). 
The package includes routines to exposure and vignetting correct images, as
well as to model and subtract to the greatest possible extent the non-cosmic
background components.

For extended sources, the background can vary significantly over the
region of interest, requiring an independent assessment of its distribution
over the detector. Since the non-cosmic background is distributed
differently across the
detector than the cosmic background, the first has to be subtracted 
separately.
Five different contamination components have been identified 
in the non-cosmic background of PSPC observations. For a detailed
discussion we refer to Snowden et al.
(\cite{contamination}).

\section{Observations of NGC\,247}

The dwarf galaxy NGC\,247 is the one for which we have the best data.
The ROSAT data archive contains two observations of NGC\,247. The first
observation was done from December $21^{st}$, 1991 to January $6^{th}$, 
1992 with an observation time of 18240\,s and the second from June $11^{th}$ 
to $13^{th}$, 1992 with an observation time of 13970\,s. 
After exclusion of the bad time intervals ($\sim 25\%$ of the
observation time), the non-cosmic background was subtracted following
Snowden et al. (\cite{contamination}).
The cleaned data sets were then merged into a single one in order to increase 
photon count statistics.
The data were then binned into concentric rings such that four rings lie
within $D_{25}/2$. Therefore, the thickness $\Delta R$ of each ring
corresponds to $2.67$
arcmin. Obviously, the binning is a compromise between radial resolution
and statistical errors.

\begin{table}[]
  \centering    
    \begin{tabular}{|c|c|}  \hline   
       Parameter             & Value                            \\ \hline 
       Distance              & 2.53 Mpc                                             \\
       $\rm LP_{HI}$             & 11.2 kpc                                  \\
       $\rm D_{25}/2$            & 7.87 kpc (10.69 arcmin)                              \\ \hline
       Column density $\rm (n_{H})$ & $\rm 1.5 \times 10^{20} HI\,cm^{-2}$          \\\hline
    \end{tabular}
   \caption{Adopted values for the dwarf galaxy
                            NGC\,247. $\rm D_{25}/2$ is the radius at which the surface brightness
                            reaches $\rm 25^{m}~arcsec^{-2}$, $\rm
    LP_{HI}$ is the radius of the last measured HI
                            profile point (Carignan \& Puche \cite{Car}). 
The column density is taken from Fabbiano et al.(\cite{Fab}).} \label{tab3}
\end{table}

\subsection{Point source Detection}

Since we are interested in the diffuse X-ray emission, we must
eliminate all contributions due to foreground Galactic X-ray
sources, and from discrete X-ray sources associated with NGC 247. 
We follow again the procedure outlined by Snowden \& Pietsch (\cite{snowden}) 
to remove the point sources to the faintest 
possible flux limit to minimize the excess fluctuations caused by them. 
The threshold limit we used is given in
\mbox{Table~\ref{tab4} }. We carried out the point source detection
separately for the 1/4\,keV band and for the 0.5-2\,keV band. The threshold
limit guarantees that less than one spurious source will remain
within the inner part
of the field of view (25 arcmin) at a confidence level of about $3\sigma$.

\begin{table}[]
  \centering    
    \begin{tabular}{|c|c|c|c|}  \hline   
       Band      & Threshold $\rm(cts~s^{-1})$  & Excluded Area & $\sigma$   \\ \hline 
       1/4 keV   & 0.005                    & 13.41\%         & 2.93        \\
       0.5-2 keV & 0.0035                   & 12.92\%         & 2.95        \\ \hline   
    \end{tabular}
   \caption{Percentage of the excluded area within 25 arcmin
   due to point source removal.} \label{tab4}
\end{table}

\subsection{The cleaned X-ray images}

\begin{figure}[h]
  \centering \leavevmode
  \psfig{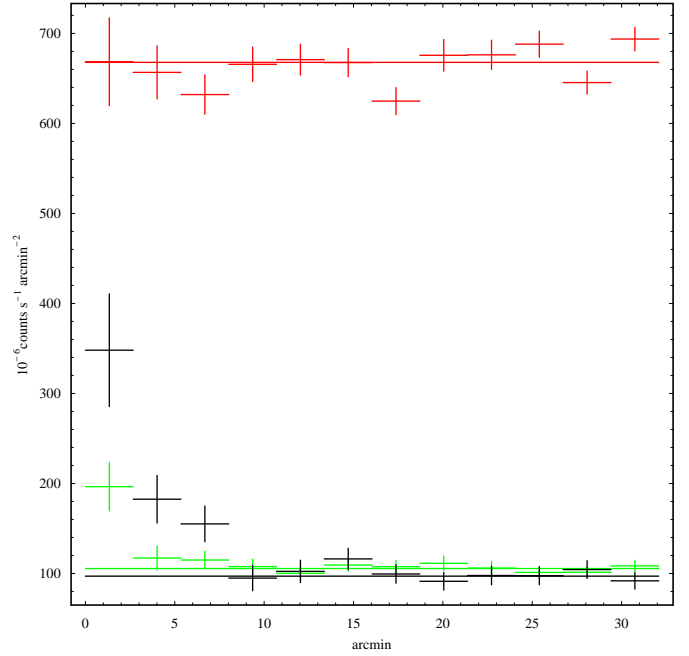}
  \vspace{-0.3cm}
  \caption{X-ray brightness radial profiles of NGC\,247 after point
           source removal for the 1/4\,keV
           band (top), the 3/4\,keV band (bottom) and the
           1.5\,keV band (middle)§. The
           horizontal lines shows the average
           intensity between the $7^{th}$ and the $12^{th}$
           annulus.}
  \label{fig5}
\end{figure}

\mbox{Figure~\ref{fig5}} shows the X-ray brightness radial profiles 
of NGC\,247 after
point source removal for the 1/4\,keV, 3/4\,keV and the 1.5\,keV band.
The horizontal lines, defined by the average X-ray
surface brightness between the $7^{th}$ and the $12^{th}$ ring, 
represent the
sum of foreground and (cosmic) background diffuse X-ray emission. 
The X-ray image in the 1.5\,keV band shows clearly that the emission 
originates from the disk of NGC\,247, which extends to about 11\,arcmin. 
We used elliptical rings for
the 1.5\,keV band, with an inclination angle of $75.4^{0}$ and a
position angle of $171.1^{0}$ as derived in Carignan \& Puche
(\cite{Car}). The distribution of the X-ray emission in the 3/4\,keV
band is less clearly associated with the disk. 
The radial profile of the 1/4\,keV band is flat from the centre up to
30\,arcmin.
The count-rate fluctuations are partly due to 
the brightest undetected point sources.

Since there exists an unresolved flux originating beyond
NGC\,247 and the HI column density of NGC\,247 is sufficiently
high over most of its inner region to absorb a significant fraction
of this flux, any emission from NGC\,247 must at least `fill' the deficit
caused by the absorbed flux in order to be observed as an
enhancement. Thus, even the flat distribution of the X-ray brightness
profile of NGC\,247 is already a clear indication for
an emission from the galaxy. Hence, we must correct the surface brightness
of the image for the absorption of the extragalactic flux.

Moreover, the increasing size of the HI column density towards the centre 
reduces the point-source detection sensitivity, and thus
more undetected X-ray sources contribute to the background. 

The contribution of undetected sources of the extragalactic background can
be calculated using the $\log N-\log S$ relation

\begin{equation}
  N(S) = \left\lbrace \begin{array}{ll}
                (285.3 \pm 24.6)~ S^{-(2.72 \pm 0.27)} & \mbox{if~$S \geq 2.66$}\\
                (116 \pm 10)~~~ S^{-(1.80 \pm 0.08)}   & \mbox{otherwise}  
                \end{array}
       \right.
\end{equation} 
as derived from a deep ROSAT survey by Hasinger et al.~(\cite{Has}). 
S denotes the flux in units of $\rm 10^{-14}\,erg~cm^{-2}s^{-1}$.  
The energy spectrum of sources with fluxes in the range 
$\rm (0.25 - 4) \times 10^{-14}\,erg~cm^{-2}s^{-1}$ is given by 
\begin{equation}
  (7.8 \pm 0.3)~E^{-(0.96 \pm 0.11)}~~~~~~[\rm keV~cm^{-2}s^{-1}sr^{-1}keV^{-1}]~,
\end{equation}
where $E$ is the energy in keV. Since the absolute
normalization for the $\log N-\log S$ relation is still uncertain, 
we will use it in a way such that only the ratio is relevant. 

The remaining count-rate $\rm I_{eg}^{i}$ of
extragalactic sources, due to sources having fluxes below the flux
detection limit $\rm S_{lim}^{i}$ (see below) 
is given by
\begin{equation}
  I_{\rm eg}^{i}=I_{\rm egobs}^{i} \frac{\int_{0}^{S_{\rm lim}^{i}}S~N(S) dS}
                                {\int_{0.25}^{4} S~N(S) dS }~,
\end{equation}
where $I_{\rm egobs}^{i}$ is the observed extragalactic count-rate in the
$i^{\rm th}$~ring of sources with a flux in the range $\rm (0.25-4) \times
10^{-14}\,erg~cm^{-2}s^{-1}$. $\rm I_{egobs}^{i}$ in units of $\rm
counts~s^{-1}arcmin^{-2}$ is given by
\begin{equation}
I_{\rm egobs}^{i} = \frac{1}{A_i}\left(\frac{2 \pi}{360\times 60}\right)^2
\int_{\rm A_{i}}f_{n}(n_{\rm HI}({\bf x}))\,d{\rm A_i} 
\end{equation}
with $A_i$ the surface of the $i^{\rm th}$ ring and
\begin{eqnarray}\label{eq4}
& &  f_{n}(n_{\rm HI}({\bf x}))=\int_{0}^{\infty}e^{-n_{HI}({\bf x}) \sigma (E)}
                           A_{\rm eff}^{n}(E)~7.8~E^{-1.96} dE\nonumber \\
\end{eqnarray}
is the observed extragalactic count-rate at position ${\bf x}$ in the
$n-$band, where $n$ labels the $1/4\,$keV, $3/4\,$keV and $1.5\,$keV band,
respectively. The column
density $\rm n_{\rm HI}$ includes both the column density of the Milky
Way ($\rm 1.5 \times 10^{20}~HI~cm^{-2}$ (Fabbiano et al. 
\cite{Fab})) and 
the column
density of NGC\,247 at position ${\bf x}$, as calculated from Carignan \& Puche
(\cite{Car}).
The source detection limit $\rm I_{Threshold}$ of our analysis (see
Table~\ref{tab4}) is then given via the relation
\begin{equation}\label{eq5}   
  I_{\rm Threshold}=c^{i} \int_{\rm A_{i}}f_{n}(n_{\rm HI}({\bf x}))
\,d{\rm A_i}
\end{equation}
in units of $\rm counts~s^{-1}$. The flux detection limit $\rm S_{lim}^{i}$
in keV is
\begin{equation}\label{eq6}
  S_{lim}^{i}=c^{i} \int_{0.5}^{2} 7.8~E^{-0.96} dE~. 
\end{equation}
Hence, using Eqs.~(\ref{eq5}) and (\ref{eq6}) we get 
\begin{eqnarray}
& &  S_{\rm lim}^{i}=1.602 \times 10^5  \frac{{\rm A_i}\,I_{\rm Threshold}~\int_{0.5}^{2} 7.8~E^{-0.96} dE}
{\int_{\rm A_{i}}f_{n}(n_{\rm HI}({\bf x}))\,d{\rm A_i}}
\nonumber \\
\end{eqnarray}
in units of $\rm 10^{-14}\,erg~cm^{-2}s^{-1}$. The extragalactic background
and Milky Way corrected count-rate $\rm I_{MW}^{i}$ in the $i^{\rm th}$ ring 
is now given by
\begin{equation}\label{mwcr}
  I_{MW}^{i}=\frac{(I_{\rm obs}^{i}+I_{\rm eg}^{i})-
         (I_{\rm obs}^{\rm Base}+I_{\rm eg}^{\rm Base})}{\tau_{f}}\,,
\end{equation}
where the base values, defined by the outermost rings ($7-12$), are treated
correspondingly.
$\tau_{f}$ is the band averaged absorption coefficient of our
own Galaxy computed using a Raymond-Smith emission
model (Raymond \& Smith \cite{Ray}) and a column density of $\rm 1.5 \times
10^{20}~HI~cm^{-2}$. We assume that the halo emission is
not absorbed by the disk material of NGC\,247.

A detailed error calculation should include the uncertainties of the 
foreground column density, the column density of NGC\,247, 
the $\log N-\log S$ relation, the extragalactic source spectrum, 
the Raymond-Smith model, including
the uncertainty in the temperature and the response matrix.
Since this calculation is in principle possible but sophisticated, we
adopt the formal error to be two times the count-rate uncertainty hence,
$\Delta I_{\rm MW}^i=2 \times \Delta I_{\rm obs}^i~/\tau_f$.

\begin{figure}[h]
  \centering \leavevmode
  \psfig{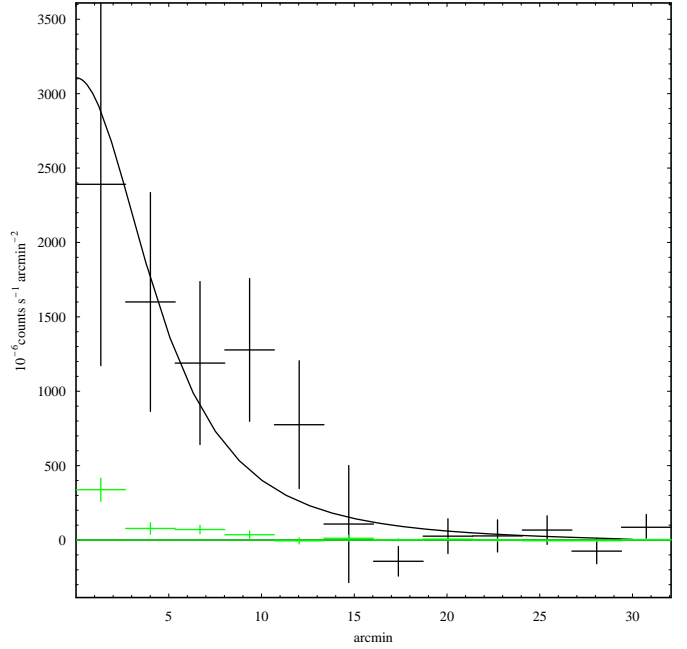}
  \vspace{-0.3cm}
\caption{Corrected count-rate in the 1/4\,keV band and the 3/4\,keV
band. For the 1/4\,keV band the best fit for $\beta=2/3$ is
superimposed.} \label{fig6}
\end{figure}


\begin{table}[]
 \centering    
  \begin{tabular}{|c|c|c|c|}  \hline
    annulus i   &  $\rm I_{MW}^i$ (1/4) & $\rm I_{MW}^i$
  (3/4) & $\rm I_{obs}$ (1.5) \\  \hline
       1        &$\rm 2391\pm 1219$ &$\rm 338\pm 78.0 $  &$\rm 251\pm 62.6 $ \\
       2        &$\rm 1600\pm 736 $ &$\rm 78.1\pm 38.6 $ &$\rm 85.5\pm 26.5$ \\
       3        &$\rm 1189\pm 548 $ &$\rm 70.6\pm 28.4 $ &$\rm 58.0\pm 19.9$ \\
       4        &$\rm 1278\pm 480 $ &$\rm 35.8\pm 24.1 $ &$\rm -2.2\pm 14.0$ \\
       5        &$\rm 775\pm 430 $  &$\rm -5.3\pm 20.8 $ &$\rm 5.3 \pm 12.5$ \\
       6        &$\rm 108\pm 100 $  &$\rm 11.0\pm 19.7 $ &$\rm 19.1\pm 11.9$ \\ \hline
 \end{tabular}
 \caption{Average 1/4\,keV and 3/4\,keV diffuse emission after correction
           for the absorption of the more distant extragalactic flux in units of
           $\rm 10^{-6}counts~s^{-1}arcmin^{-2}$. Since the absorption
           correction in the 1.5\,keV band is negligible the average
  1.5\,keV
           emission is not corrected but only background subtracted.}
  \label{tab5}
\end{table}

The corresponding radial profiles for the 1/4\,keV and the 3/4\,keV band
are shown in Fig.~\ref{fig6} and the values are given in
Table~\ref{tab5}. For the 1.5\,keV band the absorption due to gas
associated with NGC\,247 is negligible. Adopting the values given in 
Table~\ref{tab3} the X-ray luminosity corresponding to the observed
count rates in the different bands amounts to $L_X=1.1\times 10^{39}\,{\rm erg/s}$, 
$L_X=4.8\times 10^{38}\,{\rm erg/s}$ and
$L_X=2.3\times 10^{38}\,{\rm erg/s}$ in the $1/4\,$keV, $3/4\,$keV and $1.5\,$keV
band, respectively.

\section{Spectral Analysis}

The ratio of the count rates in the different energy bands contains
information about the temperature of the emitting gas.
Assuming a Raymond-Smith spectral model (Raymond \& Smith \cite{Ray})
with cosmic abundances (Allen \cite{allen}), the observed ratio as
derived from the count rates (Eq.~(\ref{mwcr})) must equal the theoretical
count rate ratio given by:
\begin{equation}
\eta = \frac{\int_{0}^{\infty}A_{\rm eff}^{n}
R(E,T)\frac{1}{E}dE}{\int_{0}^{\infty}A_{\rm eff}^{n'} 
R(E,T)\frac{1}{E}dE}\,.
\end{equation} 
For the emission below $3/4$\,keV we obtain $\log T = 6.1^{+0.1}_{-0.1}$.
The higher energy band is not included in the fit since its emission is
dominated by the contribution from the disk.

The temperature obtained from the X-ray data tends to be too large due
to the flux from the hotter plasma associated with the disk. The formal
error is expected to be too small since at a temperature of
$\log T = 6.1$ most of the emission lies outside the ROSAT bands hence,
the spectral fit has to be done using only the tail of the emission
function.   

Alternatively, we can use the rotation curve of NGC\,247
(Burlak \cite{burlak}) to obtain the temperature at the last measured HI
profile point ($11.2\,$kpc). Doing this we obtain $\log T = 5.6$, which
is in reasonable agreement with the above value deduced from X-ray data.
Both values have their own merits. Using the higher value $\log T =
6.1$ allows to determine --within the model-- the total mass
and the mass of the hot gas using only X-ray data, thereby we independently
check the total mass as obtained from the rotation curve. 
However, the error of the result is quite large. Using the lower value 
$\log T = 5.6$, obtained from the measured rotation curve, we mix the two 
methods and thus we can no longer compare with the HI-measurements, but 
the error is possibly smaller.
We will give in the following all results for both
temperatures $\log T = 5.6$ and, in brackets, $\log T = 6.1$.  

\section{Mass Determination}

\subsection{The Hot Emitting Gas}

Most of the flux originating from NGC\,247 is observed in the $1/4$\,keV band. We
determine the mass of the emitting hot gas making the following assumptions:
the gas temperature is about $\log T = 5.6$ ($\log T = 6.1$); the spectral 
emissivity is given by a Raymond-Smith model and the density profile is 
described by a $\beta$-model (see Sect.~2) with central gas 
density $\rho_{\circ}$ and core radius $a_g$.
Due to the small amount of data, we reduce the degrees of freedom by fixing  
$\beta $. 
The measured radial profile in the $1/4$\,keV band can then be used
to determine the remaining two parameters $\rho_{\circ}$ and $a_g$.
With these values we determine the integrated
gas mass within the last measured HI profile point ($r=11.2\,$kpc). 
The results for three different values of $\beta$ are given in 
Table~\ref{tab5b}.

Read et al. (\cite{read}) have analysed a sample of 17 nearby
galaxies among them NGC\,247. They found a hot gas mass in NGC\,247 of
only $3.1\times 10^6\sqrt{\eta}\,{\rm M_{\odot}}$ with $\eta$ the
filling factor, which is nearly two order of magnitudes below the average
value of their sample. Beside the different analysis technique, we believe 
that the main reason for the large discrepancy between their and our result 
is due to the different treatement of the internal absorption in NGC\,247.

\subsection{The Gravitating Mass}

The observed photons originate from a hot gas at a temperature
of $\approx 10^5-10^6$\,K as it cools due to Bremsstrahlung, and
especially in the lower temperature range more important, due
to line cooling. The amount 
of hot gas in a DG is not large enough to have a cooling time
comparable with a Hubble time, in fact for NGC\,247 we expect a cooling 
time of about $5\times 10^8$ years (Read et al.
\cite{read}).
Therefore, the gas must have been reheated either by heat sources, or
e.g. by adiabatic compression as the result of a flow through the
galaxy's potential. Obviously, both effects can combine. 
The main heat sources for the gas are supernovae. However,
the heating and cooling rates for the gas do not balance at every
radius. Since the cooling rate per ion is proportional to the local
electron density, cooling will generally dominate over heating at
small radii, where the density is high. Thus, the gas in the central
regions will steadily cool and flow into the centre of the galaxy,
where it presumably forms gas clouds and stars. However, the flow 
velocity is usually much less than the sound speed, such that approximately
hydrostatic equilibrium is 
maintained in the gas (Binney \& Tremaine \cite{Bin}).
\begin{figure}[h]
   \centering \leavevmode
\epsfig{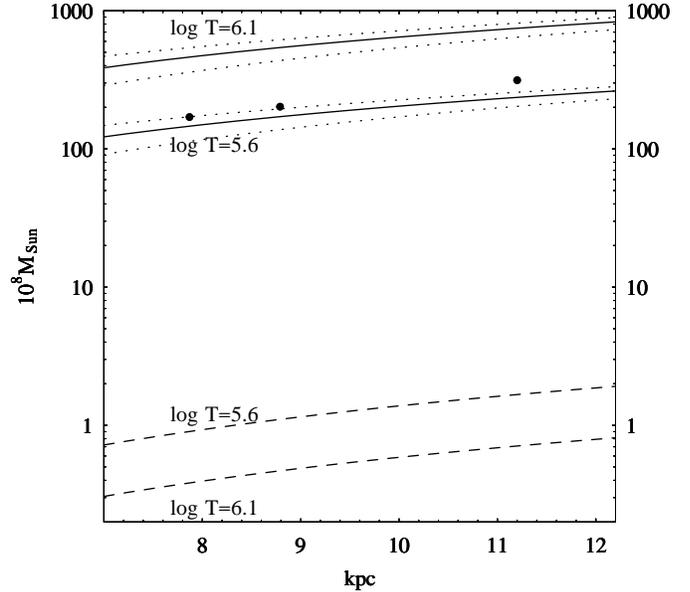}
  \vspace{-0.3cm}
  \caption{Profile of the integrated total mass of
         NGC\,247 (full lines) between 7 and 12.2 kpc for both
         $\log T = 5.6$ and $\log T = 6.1$ adopting $\beta=2/3$. The dotted lines show
         the variation of the total mass due to a $1\sigma$ error in the
         core-radius $a_g$. The dashed lines show the profile of the
         integrated hot gas mass. The three points show the total mass from 
	 Table 3 of Burlak (\protect\cite{burlak}).}
    \label{fig11}
\end{figure}

Assuming an ideal gas in hydrostatic equilibrium,
the total mass $M(r)$ within the radius $r$ is given by
Eq.~(\ref{M(r)}).
Using the values for the parameters $\rm\rho_{\circ}$ and $a_g$ from
Table~\ref{tab5b}, we can calculate the total mass of
NGC\,247 within $r=11.2\,$kpc.  
The results are given in Table~\ref{tab5b}.
\mbox{Figure~\ref{fig11}} shows the profile of the 
integrated total mass for $\beta=2/3$ (full lines). The dotted lines give the variation due
to a $1\sigma$ error in $a_g$.
The three points are the values taken from Table 3 of Burlak~(\cite{burlak}).
At $r=11.2\,$kpc he finds $M=3.1\times 10^{10}\,{\rm M_{\odot}}$, where we
refer to his dark halo model. We see, that given the uncertainties in the
data, the agreement is still reasonable within a factor of two.
\begin{table*}[]
 \centering 
\renewcommand{\arraystretch}{1.2}   
  \begin{tabular}{|r|c|c|c|c|c|}  \hline
    & $\rho_{\circ}\,[10^{-4}\,{\rm M_{\odot}/pc^3}]$ & $a_g\,[{\rm kpc}]$ &
    $M_{hg}\,[10^{8}\,{\rm M_{\odot}}]$ & $M\,[10^{10}\,{\rm M_{\odot}}]$ & $\chi^2_{red}$\\  \hline
    $\log\,T=5.6$ & $$ & $$ & $$ & $$ & \\
    $\beta=1/2$ & $1.4^{+0.9}_{-0.5}$ & $2.6^{+1.3}_{-1.0}$ & $1.6^{+0.9}_{-0.7}$ & $1.9^{+0.1}_{-0.1}$ & $2.43$ \\
    $\beta=2/3$ & $1.2^{+0.7}_{-0.4}$ & $4.4^{+2.2}_{-1.6}$ & $1.7^{+1.1}_{-0.8}$ & $2.4^{+0.2}_{-0.3}$ & $2.07$ \\
    $\beta=1$ & $1.0^{+0.5}_{-0.4}$ & $7.1^{+3.5}_{-2.7}$ & $1.8^{+1.1}_{-1.0}$ & $2.9^{+0.6}_{-0.8}$ & $1.74$ \\
    $\log\,T=6.1$ & $$ & $$ & $$ & $$ & $$ \\
    $\beta=1/2$ & $0.6^{+0.4}_{-0.2}$ & $2.6^{+1.3}_{-1.0}$ & $0.7^{+0.4}_{-0.3}$ & $6.1^{+0.2}_{-0.4}$ & $2.43$ \\
    $\beta=2/3$ & $0.5^{+0.3}_{-0.2}$ & $4.4^{+2.2}_{-1.6}$ & $0.7^{+0.5}_{-0.4}$ & $7.5^{+0.7}_{-1.0}$ & $2.07$ \\
    $\beta=1$ & $0.4^{+0.2}_{-0.2}$ & $7.1^{+3.5}_{-2.7}$ & $0.8^{+0.5}_{-0.4}$ & $9.2^{+2.0}_{-2.4}$ & $1.74$ \\
\hline
 \end{tabular}
 \caption{The physical parameters of the hot diffuse gas and the total mass
          within $11.2\,$kpc for $\log\,{\rm T}=5.6$ and $\log\,{\rm T}=6.1$
	  and several values of $\beta$. The errors for $M_{hg}$ and $M$ are
          mainly due to the error in the core radius $a_g$.}
  \label{tab5b}
\end{table*}

\section{Discussion and conclusions}

We determined the mass of the hot emitting gas and the total mass of the 
dwarf galaxy NGC\,247 using ROSAT PSPC data. From the X-ray data we obtain 
$M_{hg}=0.7\times 10^{8}M_{\odot}$ for the hot gas mass and
$M=7.5\times 10^{10}M_{\odot}$ for the total mass adopting $\beta=2/3$,
which seems to be favoured by the measured rotation curve. 
Unlike, as suggested by the results published so far, NGC\,247 is
not an exceptional object with a hot gas content about two orders
of magnitude below the group average. The result obtained in this analysis
agrees well with what is found for other Sculptor members. Although the 
quality of the X-ray data is less good as compared to e.g. elliptical 
galaxies, the mass determination agrees well with the value found by 
independent methods. The relative amount of hot gas is comparable to what 
one finds in more massive galaxies (DePaolis et al. \cite{strafa}). For the 
mass to luminosity ratio as derived from the X-ray data, we obtain 
$\log(L_X/M) = 28.4$ which agrees well with values found for other Sculptor 
members (Read \& Pietsch \cite{rpi}, Schlegel et al. \cite{schl}).

The hot emitting gas seems to form a halo around the galactic center which
is more concentrated towards the center than the dark halo. In fact, we find
a core radius of $a_g=4.4\,$kpc for the hot gas, whereas Puche \& Carignan
(\cite{puche}) favour an extended dark halo with a core radius of
$24.2\,$kpc. Burlak's dark halo model (Burlak \cite{burlak}) assumes a core
radius of $a=6.9\,$kpc. The smaller core radius of the hot gas found by
the X-ray analysis could indeed indicate a flow of gas towards the center.


As one can see in Fig.~\ref{fig5}, the brightness profile of NGC\,247
in the $3/4\,$keV and $1.5\,$keV band shows the presence of 
a ``hump'' about $15\,$arcmin from the center. Similar observations 
are known from M\,101 and M\,83. If MACHOs are low-mass
stars or even brown dwarfs, we expect them to be X-ray active (De Paolis et
al., \cite{dijr98}) with an X-ray luminosity of about
$10^{27}-10^{28}\,$erg/s particularly during their early stages (Neuh\"auser
\& Comer${\rm \acute o}$n, \cite{neuhauser}). Using
similar arguments as in De Paolis et al. (\cite{dijr98}), it can be shown
that this ``hump'' could be explained by the X-ray emission of dark clusters
of MACHOs, which could make up to half of the dark matter in NGC\,247. 
Therefore, NGC\,247 might offer the unique possibility to observe
how a halo of MACHOs actually forms.



\begin{acknowledgements} 
The authors acknowledge helpful discussions with Alexis Finoguenov, 
Christine Jones, and Steve Snowden. Special thanks go to Claude Carignan
for providing the NGC\,247 HI-map in digital form, and to Steve Snowden
for providing the detector on-axis response curves. We would like to thank
the referee for his helpful and clarifying comments.

This work is partially supported by the Swiss
National Science Foundation.
\end{acknowledgements}

\end{document}